\documentclass[aps,prd,preprintnumbers,showpacs]{revtex4}
\usepackage{graphicx}

\begin{document}

%
%

\eprint{Nisho-1-2026}
\title{Power of Axion Microwave Absorbed by Quantum Hall State in Haloscope}
\author{Aiichi Iwazaki}
\affiliation{Faculty of International Politics and Economics, Nishogakusha University,\\ 
6-16 3-bantyo Chiyoda Tokyo 102-8336, Japan }   
\date{July. 24, 2026}
\begin{abstract}
We propose a new method for detecting dark matter axions using a resonant cavity coupled to a two-dimensional electron system in the quantum Hall regime. When the cavity is tuned to the axion frequency, the axion-induced electromagnetic field is resonantly enhanced and drives a transverse Hall current in the quantum Hall system. On a quantum Hall plateau, the longitudinal dissipative response is strongly suppressed, $\mathrm{Re}(\sigma_{xx})\simeq0$, while the Hall conductivity remains finite and quantized, $\mathrm{Re}(\sigma_{xy})=\nu e^2/h$. Consequently, the Hall current is essentially nondissipative and introduces only a small additional loss to the cavity, allowing the loaded quality factor to approach the unloaded value, $Q_L\simeq Q_0$. The resulting Hall current is therefore enhanced by the large cavity quality factor, $I_H\propto\mathrm{Re}(\sigma_{xy})E\propto Q_L$. For a 2D electron density of $3\times10^{11}\mathrm{cm}^{-2}$, filling factor $\nu=2$, and $Q_L\sim10^5$--$10^6$, we estimate a Hall current of order $I_H\sim10^{-13}$--$10^{-12}\mathrm{A}$ for an axion mass $m_a\sim10^{-5}\mathrm{eV}$ and a magnetic field $\sim 5\times 10^4\mathrm{G}$. The current is more enhanced by increasing
the size of 2D electrons keeping the strength of magnetic field. 
Under the assumed thermal-noise level and readout conditions, the estimated Hall-current signal can achieve a signal-to-noise ratio greater than unity for an observation time of order $100\mathrm{s}$. The proposed method exploits the unique combination of a finite, quantized transverse response and a strongly suppressed longitudinal dissipation in the quantum Hall state, providing an alternative to conventional metallic-antenna detection in axion haloscopes.

\end{abstract}
\hspace*{0.3cm}

\hspace*{1cm}

\maketitle


\section{introduction}

One of the most important problems in particle physics is to clarify the nature of dark matter in the Universe. Among the various dark matter candidates, the axion \cite{axion1,axion2,axion3} is particularly promising, since it can constitute the dark matter while providing a natural solution to the strong CP problem. Despite extensive experimental efforts to detect axions (see, e.g., the references in \cite{review}), however, no axion signal has yet been observed.

In this paper, we propose a novel method for axion detection based on a quantum Hall state \cite{girvin} incorporated into a microwave haloscope. Haloscopes have been widely employed in axion searches using strong magnetic fields \cite{sikivie,admx1,admx2}. However, the sensitivity to axions with masses $m_a \gtrsim 10^{-5}$ eV becomes increasingly challenging because the volume of a resonant cavity decreases rapidly with increasing axion mass. In conventional haloscope experiments, the axion-induced microwave radiation is extracted using a metallic antenna. The electromagnetic dissipation associated with the antenna reduces the loaded quality factor $Q_L$ of the cavity and, consequently, suppresses the resonant enhancement of the axion-induced signal.

Microwave radiation generated by axions in a resonant cavity is amplified by the cavity quality factor $Q_L$, and the amplified signal is conventionally detected using a metallic antenna. We propose instead to use a semiconductor sample exhibiting the quantum Hall effect as the detector. In a quantum Hall state, the longitudinal conductivity $\sigma_{xx}$ is strongly suppressed, while a transverse Hall current can flow in response to an electric field. Since the dissipative component of the response is governed by $\mathrm{Re}(\sigma_{xx})$, a quantum Hall state with $\mathrm{Re}(\sigma_{xx})\simeq 0$ introduces substantially less microwave loss than a conventional metallic antenna. Consequently, the cavity can retain a large loaded quality factor, allowing the axion-induced microwave field to be resonantly enhanced.

\begin{figure}[htp]
\centering
\includegraphics[width=0.6\hsize]{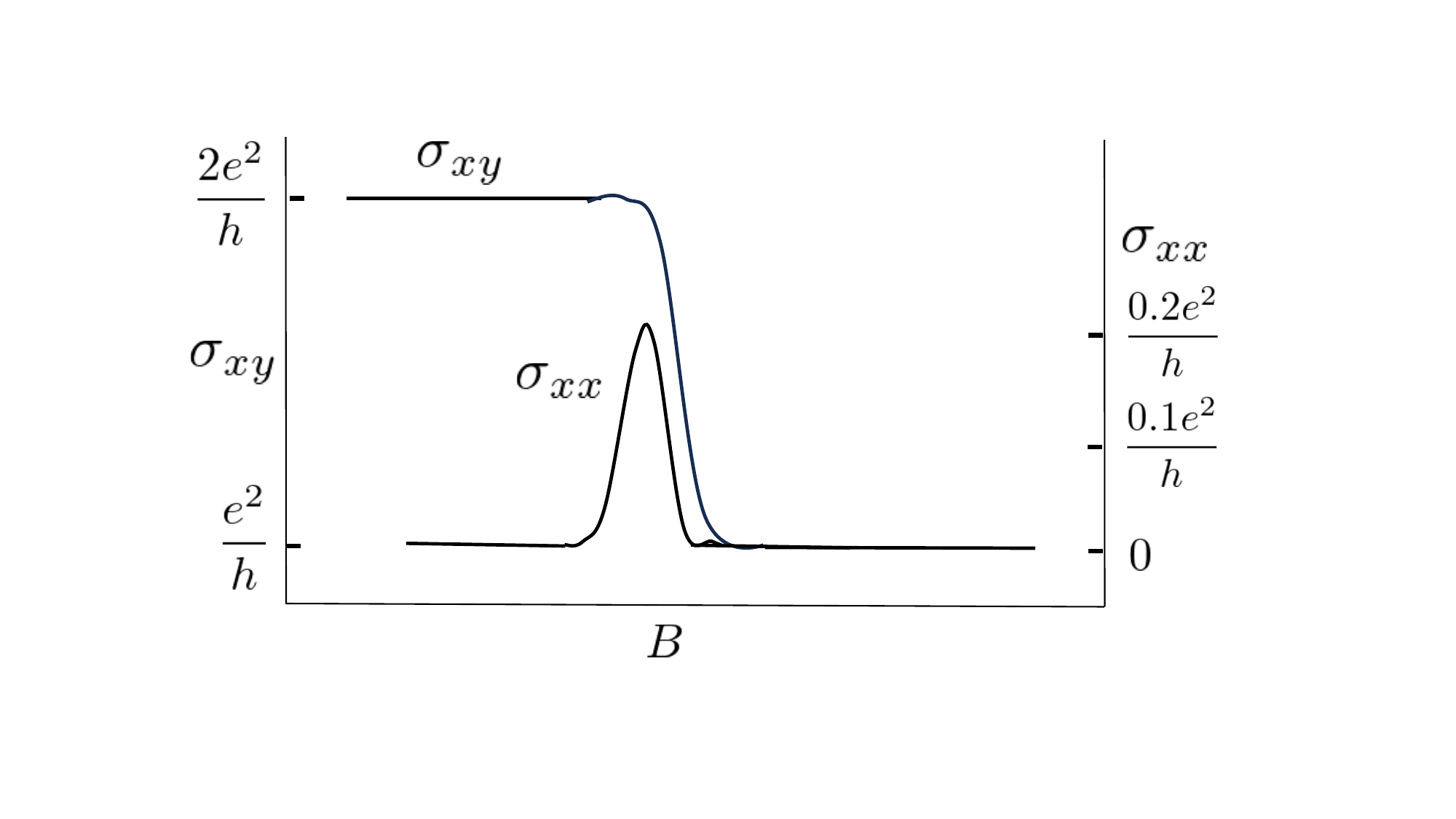}
\caption{Schematically figure of $\sigma_{xy}$ plateau and $\sigma_{xx}$ in the transition region between plateaus} 
\label{aaa}
\end{figure}

In quantum Hall systems, the Hall conductivity $\sigma_{xy}$ exhibits quantized plateaus as a function of magnetic field and takes the values
\begin{equation}
\sigma_{xy}=\frac{\nu e^2}{h}=\frac{\nu e^2}{2\pi},
\end{equation}
where $h$ is Planck's constant and $\nu$ is an integer ($\nu\geq1$) or a fractional number such as $\nu=2/3,,1/3,\ldots$. 
It is called as filling factor. The center of the plateau is realized at the magnetic field $B$ such as $2\pi\rho_e/eB=\nu$,
where $\rho_e$ denotes number density of 2 dimensional electrons.  
On these plateaus, the longitudinal conductivity is strongly suppressed, $\sigma_{xx}\simeq0$. Thus, the quantum Hall state behaves as a nondissipative state with respect to longitudinal electrical transport and exhibits strongly suppressed microwave absorption associated with $\mathrm{Re}(\sigma_{xx})$. In contrast, $\sigma_{xx}$ becomes finite in the transition regions between adjacent Hall plateaus and reaches a maximum near the center of the transition. Typically, the maximum longitudinal conductivity is of order
\begin{equation}
\mathrm{Re}(\sigma_{xx})\sim \frac{0.1e^2}{h}
=\frac{0.1e^2}{2\pi},
\end{equation}
as illustrated in Fig.(\ref{aaa}). The transition region therefore provides a dissipative, or metallic response and can absorb microwave radiation.

We first consider the metallic quantum Hall state, for which $\mathrm{Re}(\sigma_{xx})\neq0$. For a GaAs semiconductor sample with a 2D electron area of $S=10\mathrm{cm}^2$, the corresponding quality factor of the sample is estimated to be
$Q_s\propto\sigma_{xx}^{-1}$ and of order $10^6$ for
$\mathrm{Re}(\sigma_{xx})=0.1e^2/(2\pi)$. Assuming an unloaded cavity quality factor of $Q_0=10^5$, the power dissipated by the 2D electrons in the metallic quantum Hall state under an external magnetic field $B_t$ is estimated as
\begin{equation}
P_s=\Big(g_{a\gamma\gamma}B_t\Big)^2\frac{\rho_d}{m_a}VC\frac{Q_L^2}{Q_s}
\simeq 6.2\times 10^{-24}W
\Big(\frac{g_{\gamma}}{0.36}\Big)^2
\Big(\frac{B_t}{1.5\times 10^5\mathrm{G}}\Big)^2
\Big(\frac{V}{7.2\mathrm{l}}\Big)
\Big(\frac{C}{0.6}\Big)
\Big(\frac{\rho_d}{0.45\mathrm{GeVcm^{-3}}}\Big),
\end{equation}
where the loaded quality factor is $Q_L=9.2\times10^4$, and
\begin{equation}
Q_s=1.1\times10^6
\Big(\frac{10^{-5}\mathrm{eV}}{m_a}\Big)
\Big(\frac{10\mathrm{cm}^2}{S}\Big)
\Big(\frac{0.2e^2/h}{\mathrm{Re}(\sigma_{xx})}\Big).
\end{equation}
Here, $V$ and $C$ denote the cavity volume and form factor, respectively. The parameter $g_\gamma$ is model dependent, with $g_\gamma(\mathrm{DFSZ})\simeq0.36$ for the DFSZ model \cite{dfsz,dfsz1} and $g_\gamma(\mathrm{KSVZ})\simeq-0.96$ for the KSVZ model \cite{ksvz,ksvz1}. $\rho_d$ denotes density of axion dark matter.
At resonance, the cavity volume is
\begin{equation}
V=\pi R^2l_{\text{cav}}
\simeq \pi\Big(\frac{2.4}{m_a}\Big)^2
\times10^2\mathrm{cm}
\Big(\frac{l_{\text{cav}}}{10^2\mathrm{cm}}\Big)
\simeq7.2\times10^3\mathrm{cm}^3
\equiv7.2\mathrm{l}
\Big(\frac{10^{-5}\mathrm{eV}}{m_a}\Big),
\end{equation}
where $l_{\text{cav}}$ is the cavity length. This is the power dissipated by the metallic quantum Hall state and is essentially analogous to the power dissipated by a conventional metallic antenna. Thus, the metallic quantum Hall state by itself provides no fundamental advantage over a conventional antenna.

\vspace{0.1cm}
The situation is qualitatively different when the system is operated on an insulating quantum Hall plateau at the filling factor $\nu$. In this case, $\mathrm{Re}(\sigma_{xx})\simeq 0$ is strongly suppressed, and hence the microwave power dissipated by the 2D electron system,
\begin{equation}
P_s\propto\mathrm{Re}(\sigma_{xx}),
\end{equation}
becomes very small. This remains the case even when the system is exposed to microwave radiation with frequency
$m_a/2\pi\simeq2.4$ GHz. At the same time, the Hall conductivity remains finite and quantized. The axion-induced electric field can therefore generate a Hall current
\begin{equation}
I_H=\sigma_{xy}El,
\end{equation}
where $l$ is the transverse dimension of the 2D electron system, perpendicular to the electric field $\vec E$. We obtain
\begin{equation}
I_H\simeq1.66\times10^{-18}A
\Big(\frac{g_{\gamma}}{0.36}\Big)
\sqrt{\frac{C}{0.6}}
Q_L
\Big(\frac{\rho_e}{3\times10^{11}\mathrm{cm}^{-2}}\Big)
\Big(\frac{l}{10\mathrm{cm}}\Big),
\end{equation}
where we have adopted a 2D electron density of
$\rho_e=3\times10^{11}\mathrm{cm}^{-2}$.
The magnetic field required to realize $\nu$ quantum Hall state, for instance $\nu=1$, is approximately
\begin{equation}
B\simeq 1.2\times10^5\mathrm{G}\Big(\frac{1}{\nu}\Big)\left(\frac{\rho_e}
{3\times10^{11}\mathrm{cm}^{-2}}\right).
\end{equation}
as follows from the definition of the filling factor $\nu$,
\begin{equation}
\frac{2\pi\rho}{eB}=\nu=1.
\end{equation}

As is evident, increasing both the density $\rho_e$ and the filling factor $\nu$ by the same factor yields a larger Hall current, even though the magnetic field strength remains the same. It is a specific feature of the quantum Hall system.
Because the Hall current is nondissipative on the quantum Hall plateau, the cavity can maintain a large $Q_L$, of order 
$Q_0\sim10^5$--$10^6$. The resulting Hall current can therefore provide a detectable signal with a signal-to-noise ratio greater than unity.

As we show below, the signal-to-noise ratio of the Hall current in fractional quantum Hall state, for instance, $\nu=1/3$ is three times bigger
than the one in integer quantum Hall state $\nu=1$. However, 
the integer quantum Hall state $\nu=1$ is more easily fabricated and realized with three times
weaker magnetic field than 
fractional quantum Hall state $\nu=1/3$. Furthermore, it
has more stable longitudinal conductivity 
$\mathrm{Re}(\sigma_{xx}(m_a))\simeq 0$
against microwave with frequency
$m_a=10^{-5}\mathrm{eV}\simeq 2.4\mathrm{GHz}$.
In the paper, we propose to use integer quantum Hall states with large filling factor $\nu\ge 1$ for the axion detection.

\vspace{0.1cm}
It is important to note that the physical volume of the semiconductor sample is much smaller than the cavity volume. For example, a 2D electron system with an area of $10\mathrm{cm}^2$ and a thickness of $10$--$100\mu\mathrm{m}$ has a volume of only approximately $10^{-2}$--$10^{-1}\mathrm{cm}^3$, whereas the cavity volume considered above is
$V\simeq7.2\mathrm{l}=7.2\times10^3\mathrm{cm}^3$.
Thus, the small physical volume of the semiconductor does not determine the volume available for resonant axion conversion; rather, the cavity volume determines the resonant enhancement of the axion-induced electromagnetic field.

In summary, the use of a 2D electron system exhibiting the quantum Hall effect inside a haloscope provides an alternative to the conventional metallic antenna. We distinguish two regimes. The first is the metallic regime, characterized by a finite longitudinal conductivity $\mathrm{Re}(\sigma_{xx})\neq0$, in which the quantum Hall sample acts essentially as a dissipative antenna. The second is the insulating quantum Hall regime, characterized by $\mathrm{Re}(\sigma_{xx})\simeq0$ and a finite, quantized Hall conductivity $\sigma_{xy}$. In this regime, microwave dissipation is strongly suppressed while a Hall current can still be generated by the axion-induced electric field. This unique combination of a large transverse response and negligible longitudinal dissipation is absent in a conventional metallic antenna. We therefore propose the insulating quantum Hall state as a novel and potentially feasible approach to axion detection.

\vspace{0.1cm}

In following sections, we first review the fundamental properties of the quantum Hall state in 2D electron systems.
We then demonstrate how microwave radiation generated by dark matter axions, is absorbed by the 2D electrons or
induce dissipationless longitudinal current as well as Hall current. 
We find that detectable Hall current is obtained on the plateau of insulating quantum Hall state. 
Through out the paper,  the temperature is of the order of $1\mathrm{K}$ or less so that 
chemical potential is nearly equal to the Fermi energy $E_f$.
We only examine QCD axion.
We use physical units, Boltzmann constant $k_B=1$, light velocity $c=1$ and $\hbar=1$
( sometimes we use Planck constant $h$ ).

\section{quantum hall effect}
\label{2} 

Two-dimensional electrons in a perpendicular magnetic field $B$ form Landau levels with energies
\begin{equation}
E_n=\omega_c\left(n+\frac{1}{2}\right), \,\, n=0,1,2,\ldots,
\end{equation}
where the cyclotron frequency is
\begin{equation}
\omega_c=\frac{eB}{m^{\ast}}
\simeq
10^{-2}\,\mathrm{eV}
\left(\frac{B}{10^5\,\mathrm{G}}\right)
\end{equation}
with the effective electron mass  $m^{\ast}$ in the semiconductor. 
For example,
\begin{equation}
m^{\ast}=0.067\,m_e
\end{equation}  in GaAs with real electron mass $m_e\simeq 0.51\mathrm{MeV}$.
Throughout this paper, we consider a GaAs-based semiconductor.

The two-dimensional electron gas is formed in a quantum well of width approximately
$10\,\mathrm{nm}$ at the GaAs/AlGaAs heterointerface. At low temperatures ($T\lesssim1\,\mathrm{K}$), the electron motion perpendicular to the quantum well is frozen out because of the large confinement energy, whereas the motion parallel to the well remains free.

Each Landau level possesses a degeneracy per unit area given by
\begin{equation}
\frac{eB}{2\pi},
\end{equation}
so that, in the absence of disorder, the density of states is proportional to
\begin{equation}
\delta(E-E_n).
\end{equation}
 In real samples, a weak disorder potential $V$ broadens the Landau levels. That is, the degeneracy is lifted up.
As a result, some electronic states become localized, while others remain extended over the two-dimensional electron gas. The coexistence of localized and extended states is the essential ingredient leading to the integer quantum Hall effect\cite{girvin}. 
The integer quantum Hall effect is characterized by integer filling factor $\nu=2\pi \rho_e/eB=1,\,2\,,\,,\,,$
with electron density $\rho_e$. It implies that Landau levels up to $n$ are fully occupied and the level with $n+1$ is partially 
occupied when $n+1>\nu>n$.
On the other hand,  fractional quantum Hall effect characterized by $\nu=2/3,\, 1/3,\, \, ,\, ,$ is realized by 
Coulomb interaction between electrons, which is stronger than very weak disorder potential $V_d$.
Such states can be seen in samples with large mobility $\mu$. 

\vspace{0.1cm}

When the electron spin is taken into account, each Landau level splits into two sublevels with energies
\begin{equation}
E_{n\pm}
=
\omega_c\left(n+\frac{1}{2}\right)
\pm g\mu_B B
\end{equation}
 with the Zeeman energy $\pm g\mu_BB$ where the Bohr magneton $ \mu_B=\frac{e}{2m_e}$
and the effective $g$ factor in GaAs
$g\simeq0.44$.
The Zeeman splitting is 
\begin{equation}
g\mu_BB
\simeq
10^{-4}\,\mathrm{eV}
\left(\frac{B}{10^5\,\mathrm{G}}\right).
\end{equation}
The splitting is much lower than Landau level splitting $\omega_c\sim 10^{-2}\mathrm{eV}$.
Then, for instance, when $2>\nu>1$, Landau level with energy $E_{0-}$ is fully occupied 
and the level with energy $E_{0+}$ is partially occupied.
It is obvious that the stability of plateau specified with $\nu$ against temperature, external microwave, e.t.c. 
is high when $\nu$ is even, e.g. $\nu=2$ or $4$, because the energy gap $\omega_c$ between $E_{0+}$ ( $E_{1+}$ ) and $E_{1-}$ ( $E_{2-}$ )
is much larger than  Zeeman splitting $g\nu_BB$.

\section{metallic and insulating quantum Hall states}
\label{3}

In real samples, a disorder potential $V_d$ is inevitably present. In a two-dimensional electron system, disorder leads to localization of electronic states, and in the absence of a magnetic field the electronic states are generally localized \cite{localization}. In the presence of a strong magnetic field, however, the electronic spectrum is reorganized into Landau levels. Within each broadened Landau level, localized and extended states coexist. The extended states are responsible for electrical transport and play an essential role in the quantum Hall effect and in the absorption of microwave radiation.

\begin{figure}[htp]
\centering
\includegraphics[width=0.6\hsize]{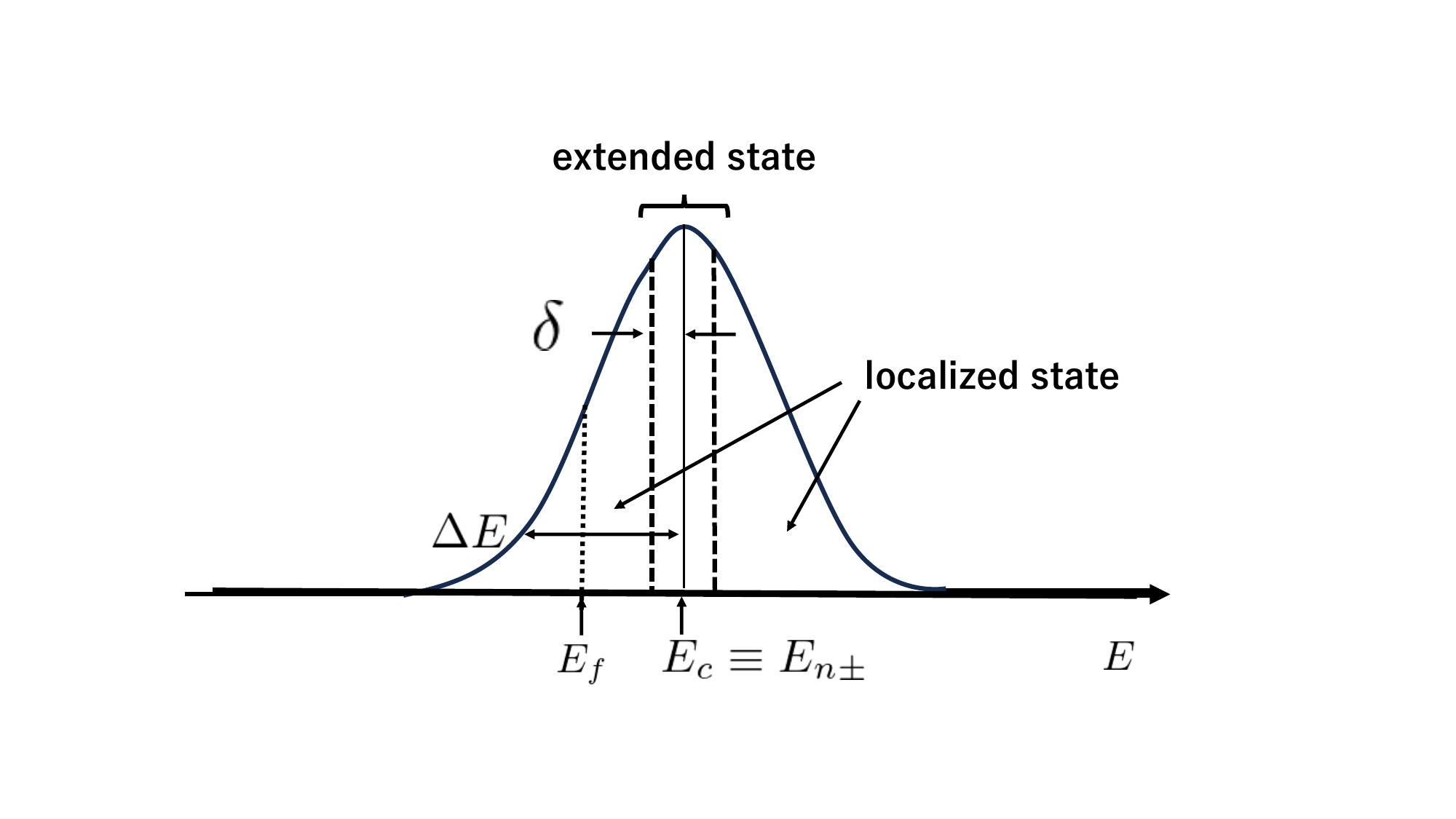}
\caption{Density of state $\rho(E)$. There are states extended over the sample and states localized.
Dotted line represents Fermi energy $E_f$}
\label{a}
\end{figure}

The broadening of the Landau levels is caused primarily by the disorder potential. Numerical calculations show that the density of states $\rho(E)$ around each Landau-level center has a finite energy width $\Delta E$. It can be schematically represented by
\begin{equation}
\rho(E)\propto
\exp\left[-\frac{(E-E_{n\pm})^2}{(\Delta E)^2}\right],
\end{equation}
as illustrated in Fig.~\ref{a}. Here, $E_{n\pm}$ denotes the center of a broadened Landau level. Since the disorder potential is much smaller than the cyclotron energy,
\begin{equation}
V_d\ll\omega_c
\simeq1.8\times10^{-2}\mathrm{eV}
\left(\frac{B}{10^5\mathrm{G}}\right),
\end{equation}
the separation between adjacent Landau levels is much larger than their disorder-induced broadening, i.e.,
$E_{n\pm}\gg\Delta E$ \cite{andouemura}.

The quantum Hall system is characterized by the filling factor
\begin{equation}
\nu\equiv\frac{2\pi\rho_e}{eB},
\end{equation}
where $\rho_e$ is the two-dimensional electron density. For $\nu<1$, only a fraction of the lowest Landau level is occupied. For $\nu>1$, the lowest Landau level is completely occupied and electrons begin to populate higher Landau levels.

As schematically shown in Fig.(\ref{a}), the states within a broadened Landau level can be divided into localized and extended states. Let $E_c\equiv E_{n\pm}$ denote the critical energy near the center of the Landau level. States satisfying
\begin{equation}
E_\alpha<E_c-\delta
\quad\text{or}\quad
E_\alpha>E_c+\delta
\end{equation}
are localized, whereas states in the critical energy interval
\begin{equation}
E_c-\delta\leq E_\alpha\leq E_c+\delta
\end{equation}
are extended \cite{aoki1,aoki2}. More precisely, the extended states are characterized by a localization length $\xi(E_\alpha)$ that becomes larger than the characteristic size of the sample. In the thermodynamic limit, the localization length diverges at the critical energy,
\begin{equation}
\xi(E_\alpha)\rightarrow\infty
\qquad\text{as}\qquad
E_\alpha\rightarrow E_c.
\end{equation}
Thus, extended states near $E_c$ can carry electrical current, whereas localized states do not contribute to dc longitudinal transport.

When the Fermi energy $E_f$ lies in the localized-state region,
\begin{equation}
E_f<E_c-\delta
\quad\text{or}\quad
E_f>E_c+\delta,
\end{equation}
the system is in a quantum Hall plateau state. Since the states at the Fermi energy are localized, the longitudinal conductivity vanishes in the dc limit,
\begin{equation}
\sigma_{xx}(0)\simeq0.
\end{equation}
At the same time, the Hall conductivity remains finite and takes a quantized value,
\begin{equation}
\sigma_{xy}=\frac{\nu e^2}{h}
=\frac{\nu e^2}{2\pi},
\end{equation}
where $\nu$ is an integer for the integer quantum Hall effect and can take fractional values for the fractional quantum Hall effect. Equivalently, the Hall resistance is quantized according to
\begin{equation}
\rho_{xy}
=\frac{h}{\nu e^2}
=\frac{2\pi}{\nu e^2}.
\end{equation}
The quantization of the Hall conductivity is a manifestation of the topological nature of the quantum Hall state \cite{topology1,topology2}.

\begin{figure}[htp]
\centering
\includegraphics[width=0.6\hsize]{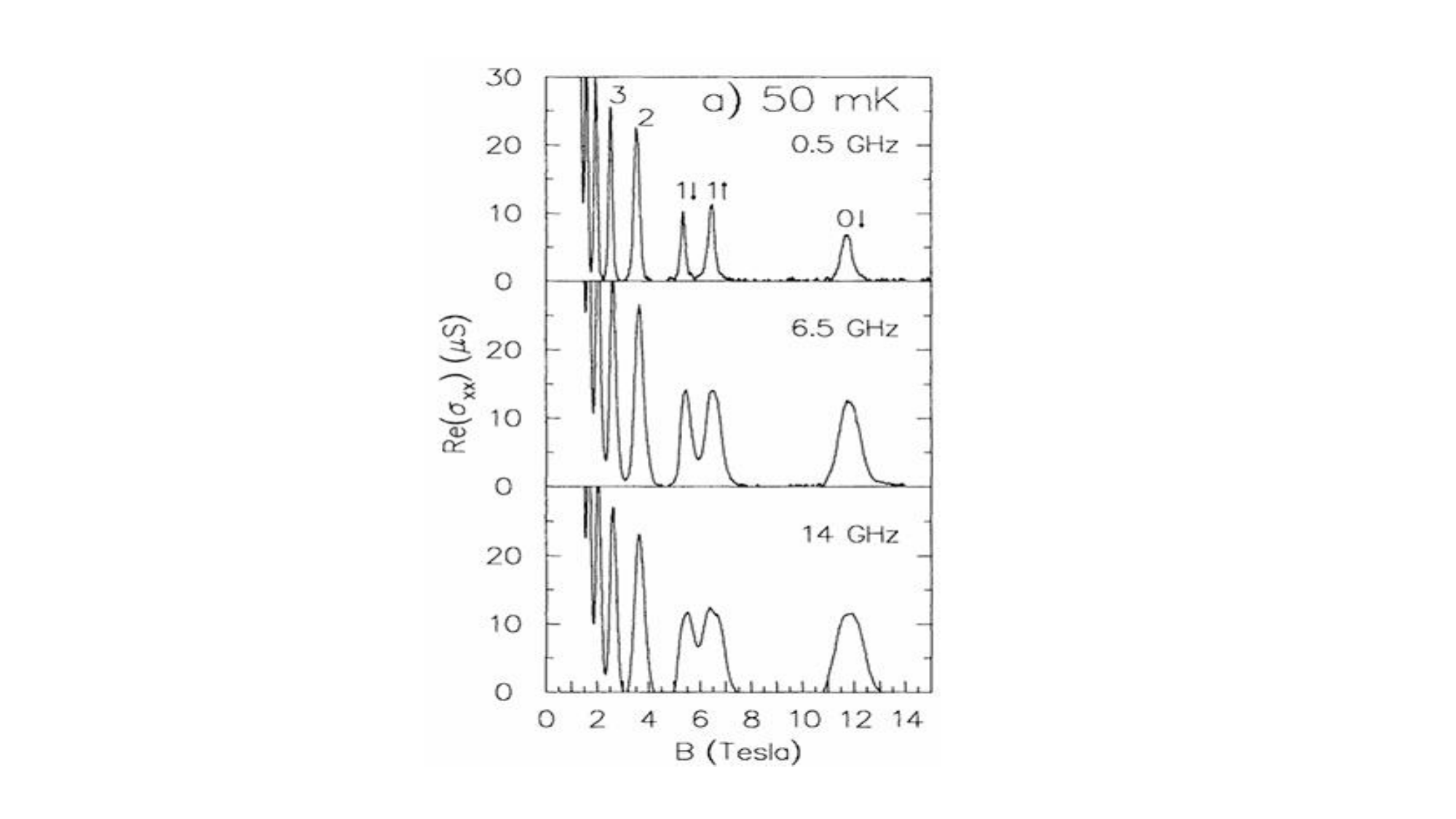}
\caption{ Measured longitudinal conductivity $\text{Re}(\sigma_{xx})$ at temperature $50\mathrm{mK}$ 
with unit $\mathrm{\mu S}\simeq \frac{0.026e^2}{h}$ 
for microwave frequencies $0.5$GHz, $6.5$GHz and $14$GHz. Ref.(\cite{sigma})} 
\label{aa}
\end{figure}

Although the longitudinal conductivity vanishes in the dc limit on a quantum Hall plateau, it is important to distinguish the dc conductivity from the finite-frequency response. For microwave radiation with frequency $\omega$, the relevant quantity governing dissipation is the real part of the longitudinal AC conductivity, $\mathrm{Re}(\sigma_{xx}(\omega))$. On a well-developed quantum Hall plateau, this quantity is strongly suppressed at sufficiently low frequencies as shown in Fig.(\ref{aa}),
\begin{equation}
\mathrm{Re}(\sigma_{xx}(\omega))\simeq0,
\end{equation}
and consequently the microwave absorption associated with the longitudinal response is strongly suppressed. The quantum Hall plateau is therefore referred to here as the \emph{insulating} or, more precisely, \emph{nondissipative} quantum Hall state. Importantly, this state is not an ordinary electrical insulator because its Hall conductivity $\sigma_{xy}$ remains finite and quantized.

In contrast, when the Fermi energy lies within the critical region,
\begin{equation}
E_c-\delta<E_f<E_c+\delta,
\end{equation}
a finite fraction of the electrons occupy extended states. These states can carry longitudinal current, and the system undergoes a plateau-to-plateau transition. In this region,
$
\sigma_{xx}\neq0,
$
and the longitudinal conductivity and Hall conductivity vary continuously with magnetic field. We refer to this regime as the \emph{metallic quantum Hall state}. The corresponding schematic behavior of $\sigma_{xy}$ and $\sigma_{xx}$ as functions of magnetic field is shown in Fig.(\ref{aaa}).

The distinction between the two regimes is particularly important for microwave absorption. For an electromagnetic field $\vec{E}$ parallel to the 2D electron system, the time-averaged dissipated power per unit area is proportional to
\begin{equation}
P_{\mathrm{abs}}\propto
\mathrm{Re}(\sigma_{xx}(\omega))|E|^2.
\end{equation}
In the metallic transition region, $\mathrm{Re}(\sigma_{xx}(\omega))$ is finite, and the 2D electrons can therefore absorb microwave radiation. The state is consequently dissipative. On the quantum Hall plateau, on the other hand, $\mathrm{Re}(\sigma_{xx}(\omega))$ is strongly suppressed, while $\sigma_{xy}$ remains finite. The microwave absorption is therefore strongly reduced, and the state behaves as a nondissipative system with respect to the longitudinal response.

In this paper, we consider both regimes. In the metallic quantum Hall state, the axion-induced microwave radiation can be absorbed by the 2D electron system through the finite longitudinal conductivity. This process is analogous to the energy dissipation in a conventional metallic antenna. In the insulating quantum Hall state, the longitudinal dissipation is strongly suppressed, allowing the resonant cavity to retain a large loaded quality factor $Q_L$. At the same time, the finite Hall conductivity allows the axion-induced electric field to generate a transverse Hall current. This combination of a large transverse electrical response and strongly suppressed longitudinal dissipation is the essential feature exploited in our proposed detection method.

Below, we discuss on the fractional quantum Hall state at filling factor
$
\nu=\frac13,
$
although the smallest accessible filling factor like $\nu=1/3$ is not necessarily optimal experimentally.
As discussed later, a smaller filling factor gives a larger Hall resistance $\rho_{xy}\propto \nu^{-1}$, 
leading to smaller thermal noise current $\propto \sqrt{1/\rho_{xy}}$
and a favorable signal-to-noise ratio for the Hall-current measurement. But, for the purpose of the 
axion detection, we propose the use of the integer quantum 
Hall state with large filling factor $\nu \ge 1$ owing to the reasons discussed in later sections.

\vspace{0.1cm}
The fractional quantum Hall state differs fundamentally from the integer quantum Hall state. It arises from strong Coulomb interactions between electrons and is realized when the disorder broadening is sufficiently small compared with the interaction energy scale. 
That is, the states are realized when disorder potential $V$ is smaller than Coulomb interaction $\sim e^2/l_B$, 
where $l_B=\sqrt{1/eB}\propto \sqrt{1/\rho_e}$ denotes
magnetic length. Thus, we need semiconductor sample with large mobility to realize fractional quantum Hall states.

The $\nu=1/3$ state can be described in terms of a strongly correlated many-body ground state, with an excitation gap determined primarily by electron-electron interactions. For typical magnetic fields of order $B\sim10^5\mathrm{G}$, the activation energy can be of order $10^{-4}\mathrm{eV}$. This energy scale is larger than the axion-induced photon energy considered here, $m_a\sim10^{-5}\mathrm{eV}$. Therefore, the axion-induced microwave field is not expected to strongly excite the fractional quantum Hall state, although the detailed finite-frequency response should ultimately be verified experimentally.

Experimentally, $\mathrm{Re}(\sigma_{xx})$ reaches a maximum near the center of the plateau-to-plateau transition. As an example, we take
\begin{equation}
\mathrm{Re}(\sigma_{xx})
\simeq\frac{0.2e^2}{h}
=\frac{0.1e^2}{\pi}
\end{equation}
near the center of the transition region. This magnitude is typical of experimentally observed longitudinal conductivities \cite{sigma,sigma2}. Measurements also indicate that the frequency dependence of $\mathrm{Re}(\sigma_{xx}(\omega))$ is relatively weak in the frequency range of interest, as shown in Fig.(\ref{aa}). Nevertheless, experimentally tuning the magnetic field precisely to the point of maximum $\mathrm{Re}(\sigma_{xx})$ may be difficult. In our proposed axion search, this experimental consideration is important because the metallic and insulating regimes play fundamentally different roles: the former provides microwave absorption, whereas the latter suppresses cavity losses while retaining a finite Hall response.

( In addition to insulating and metallic like phases, there is a superconducting-like phase \cite{joseph1,joseph2,joseph3,iwazaki}.
The phase is realized when two parallel quantum Hall layers are placed in close proximity. We have observed Josephson-like effects between such bilayer systems\cite{joseph3}. )

\vspace{1cm}

\section{axion dark matter}
\label{5}

Before discussing microwave absorption and Hall-current generation in quantum Hall systems, we briefly review the coupling of the QCD axion to the electromagnetic field \cite{axion1,axion2,axion3}.

The interaction between the axion field $a(t,\vec{x})$ and the electromagnetic field is described by the Lagrangian
\begin{equation}
\mathcal{L}_{a\gamma\gamma}=
g_{a\gamma\gamma}a(t,\vec{x}),
\vec{E}\cdot\vec{B},
\end{equation}
where $\vec{E}$ and $\vec{B}$ are the electric and magnetic fields, respectively. The axion-photon coupling constant is given by
\begin{equation}
g_{a\gamma\gamma}=
\frac{g_\gamma\alpha}{\pi f_a},
\end{equation}
where $\alpha\simeq1/137$ is the fine-structure constant, $f_a$ is the axion decay constant, and $g_\gamma$ is a model-dependent dimensionless parameter. For the QCD axion, the axion mass and decay constant are related by
\begin{equation}
m_a f_a
\simeq
6\times10^{-6}\mathrm{eV}\times10^{12}\mathrm{GeV}.
\end{equation}
Consequently,
\begin{equation}
g_{a\gamma\gamma}\propto m_a
\end{equation}
for a fixed axion model. The model-dependent coefficient is approximately
$g_\gamma\simeq0.36$ for the DFSZ model \cite{dfsz,dfsz1} and
$g_\gamma\simeq-0.96$ for the KSVZ model \cite{ksvz,ksvz1}.
In the following, we restrict our discussion to the QCD axion.

In the presence of a strong external magnetic field, the axion-photon coupling acts as an effective source for an oscillating electromagnetic field. For a spatially homogeneous axion field and a static magnetic field $\vec{B}$, the axion-induced electric field can be written schematically as
\begin{equation}
\vec{E}_a
\simeq
-g_{a\gamma\gamma}a(t)\vec{B},
\end{equation}
up to a convention-dependent sign. The oscillating electric field drives electric currents in the electromagnetic system surrounding the axion-conversion region, thereby generating an electromagnetic field at the axion frequency. In a resonant cavity, the axion-induced electromagnetic field is resonantly enhanced when the cavity mode is tuned to the axion frequency \cite{sikivie,iwazaki01}. The resulting microwave electric field can then drive a transverse Hall current in a quantum Hall system placed inside the cavity.

The axion dark matter field can be approximated as a coherently oscillating classical field,
\begin{equation}
a(t,\vec{x})=a_0
\cos\left[
\left(m_a+\frac{k^2}{2m_a}\right)t
+\vec{k}\cdot\vec{x}
+\theta_0
\right].
\end{equation}
For virialized galactic dark matter, the typical axion velocity is of order
$v\sim10^{-3}$ in natural units, and hence
\begin{equation}
|\vec{k}|\sim10^{-3}m_a,
\quad
\frac{k^2}{2m_a}\sim10^{-6}m_a.
\end{equation}
For the cavity dimensions considered below, the spatial variation of the axion field is therefore small, and we may approximate
\begin{equation}
a(t,\vec{x})
\simeq
a_0\cos(m_at+\theta_0).
\end{equation}
The corresponding axion-induced electric field oscillates at approximately the same frequency,
\begin{equation}
\vec{E}_a(t)\propto
\cos(m_at+\theta_0).
\end{equation}

The axion field has a finite coherence time and contains a narrow distribution of momenta rather than a single momentum mode. Consequently, the phase of the axion-induced electric field is not fixed indefinitely. A direct measurement of the time-averaged current,
\begin{equation}
\langle I_H\rangle,
\end{equation}
would therefore tend to vanish if the measurement time is much longer than the axion coherence time and no phase-sensitive detection scheme is employed. This issue can be avoided by measuring a phase-insensitive quantity such as the mean-square current,
\begin{equation}
\langle I_H^2\rangle,
\end{equation}
or equivalently the microwave power associated with the axion-induced signal. Since
$I_H\propto E_a$, the mean-square signal is proportional to
\begin{equation}
\langle I_H^2\rangle
\propto
\langle E_a^2\rangle.
\end{equation}
In the following, we therefore focus on the squared Hall-current signal when discussing a phase-insensitive detection of the axion-induced field.

Assuming that the local dark matter density is entirely composed of axions, the energy density is
\begin{equation}
\rho_d=
\frac{1}{2}m_a^2a_0^2
\simeq
0.45\mathrm{GeV/cm^3}.
\end{equation}
The axion field amplitude is consequently
\begin{equation}
a_0=
\frac{\sqrt{2\rho_d}}{m_a}.
\end{equation}

For the QCD axion,
$g_{a\gamma\gamma}\propto m_a$ while
$a_0\propto m_a^{-1}$.
Therefore, the product
$
g_{a\gamma\gamma}a_0
$
is approximately independent of the axion mass. This quantity determines the amplitude of the axion-induced electric field for a given external magnetic field.

The QCD axion mass is constrained by cosmological considerations and has been discussed over a broad range, approximately
\begin{equation}
10^{-6}\mathrm{eV}
\lesssim
m_a
\lesssim
10^{-3}\mathrm{eV},
\end{equation}
although the precise allowed range depends on the cosmological scenario \cite{Wil,Wil1,Wil2}. The corresponding photon frequency is
\begin{equation}
\nu_a=\frac{m_a}{2\pi}
\simeq
2.42\mathrm{GHz}
\left(\frac{m_a}{10^{-5}\mathrm{eV}}\right).
\end{equation}
Thus, axions in this mass range generate electromagnetic signals extending from the microwave to the millimeter-wave regime. In the following, we mainly consider
$
m_a=10^{-5}\mathrm{eV},
$
corresponding to a frequency of approximately $2.4\mathrm{GHz}$.

We consider a semiconductor sample containing a two-dimensional electron system with an area of $S=10\mathrm{cm}^2$ and a thickness of approximately $100\mu\mathrm{m}$ or less. Its physical volume is therefore about
\begin{equation}
V_s\simeq10^{-1}\mathrm{cm}^3,
\end{equation}
which is much smaller than the volume of the resonant cavity considered below.

For a cylindrical cavity operating in the $\mathrm{TM}_{010}$ mode, the radius required for resonance at the axion frequency is approximately
\begin{equation}
R\simeq
4.8\mathrm{cm}
\left(\frac{10^{-5}\mathrm{eV}}{m_a}\right).
\end{equation}
Taking the cavity length to be $l=100\mathrm{cm}$, its volume is
\begin{equation}
V=\pi R^2l
\simeq
7.2\times10^3\mathrm{cm}^3
\left(\frac{10^{-5}\mathrm{eV}}{m_a}\right)^2
\left(\frac{l}{100\mathrm{cm}}\right).
\end{equation}
At $m_a=10^{-5}\mathrm{eV}$, this corresponds to
$
V\simeq7.2\mathrm{l}.
$

The physical volume of the semiconductor sample is therefore much smaller than the cavity volume,
\begin{equation}
\frac{V_s}{V}
\sim10^{-5}.
\end{equation}
We consequently expect the perturbation of the $\mathrm{TM}_{010}$ cavity mode caused by the semiconductor sample to be small, provided that the sample does not introduce substantial microwave loss. The cavity form factor is taken to be
$
C=0.6,
$
which characterizes the overlap between the axion-induced source and the electric field of the resonant cavity mode.

The small physical volume of the semiconductor sample should be distinguished from the effective volume of the haloscope. The latter is determined primarily by the resonant cavity and enters the axion-conversion power through the factor $VC$. Thus, a macroscopic cavity can provide a large axion-induced electromagnetic field even when the quantum Hall sample itself occupies only a very small fraction of the cavity volume.

\section{axion microwaves in semiconductor sample}
\label{6}

We now estimate the electromagnetic field generated by axion dark matter in the resonant cavity and its interaction with a semiconductor sample containing a two-dimensional electron system. For the numerical estimates below, we use the parameters of a cylindrical haloscope. The cavity mode of interest is the $\mathrm{TM}_{010}$ mode, which is commonly used in axion haloscope experiments.

\begin{figure}[htp]
\centering
\includegraphics[width=0.6\hsize]{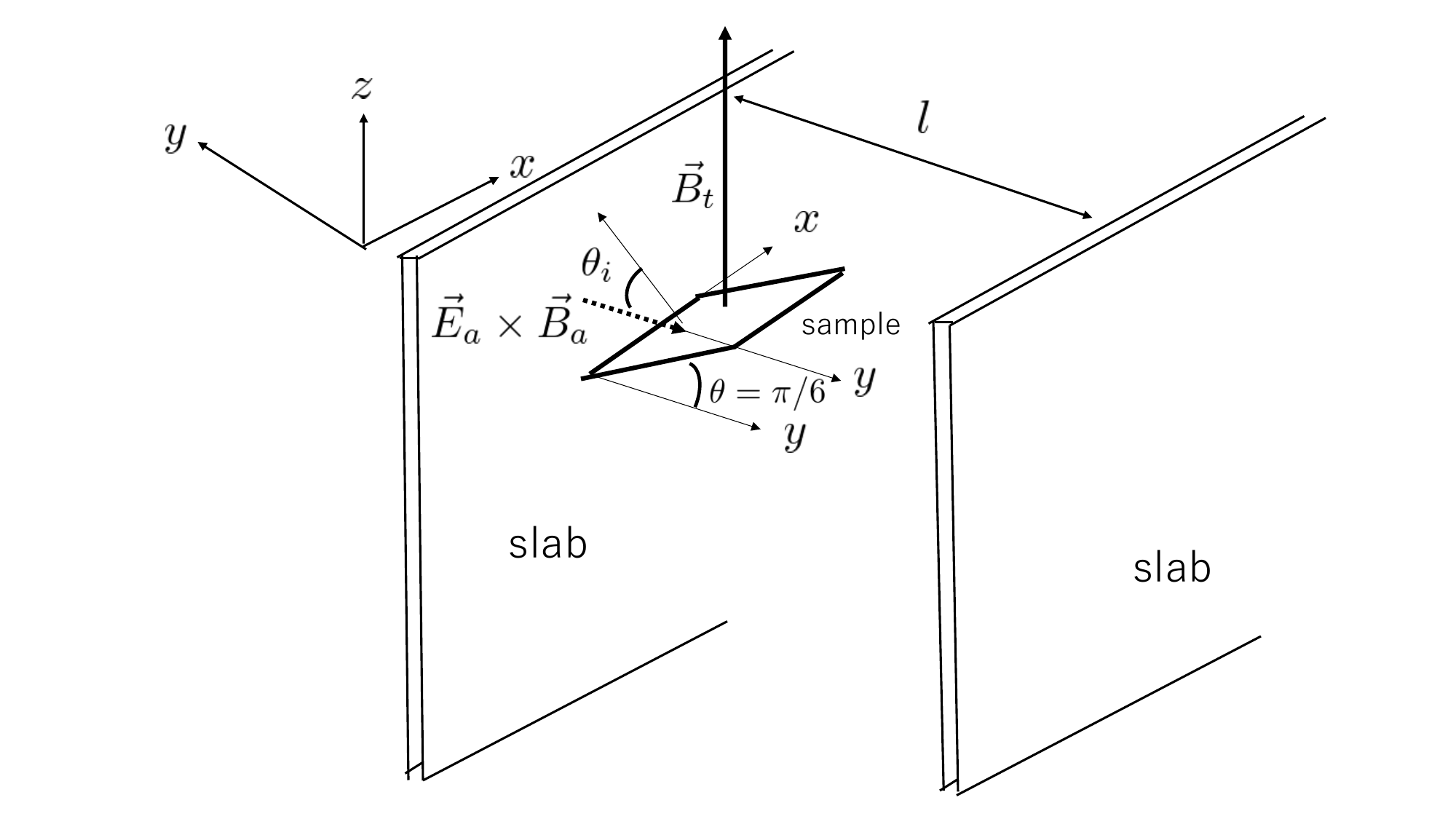}
\caption{Schematic geometry of the parallel-slab cavity. The conducting slabs are separated by a distance $l$. The semiconductor sample containing the 2D electron system is tilted by an angle $\theta=\pi/6$ with respect to the $y$ axis and is parallel to the $x$ axis.}
\label{b}
\end{figure}

For clarity, however, we first consider a simplified cavity geometry consisting of two parallel conducting slabs, as illustrated in Fig.(\ref{b}). This geometry makes the relative orientation of the microwave field, the external magnetic field, and the 2D electron layer transparent and is also useful for describing the axion-induced electromagnetic field following our previous work \cite{iwazaki01}. The results obtained from this simplified geometry will subsequently be applied to the cylindrical cavity.

Consider two parallel conducting slabs separated by a distance $l$. We apply a static external magnetic field
\begin{equation}
\vec{B}_t=(0,0,B_t)
\end{equation}
parallel to the slabs. A semiconductor sample containing a 2D electron system in a quantum Hall state is placed between the slabs. The lateral dimensions of the sample are assumed to be much smaller than the characteristic cavity length $l$.

In the presence of the external magnetic field, the axion dark matter field induces an electromagnetic wave with electric and magnetic fields that can be written as
\begin{equation}
\vec{E}_a=(0,0,E_a),
\qquad
\vec{B}_a=(B_a,0,0).
\end{equation}
The corresponding Poynting vector $\vec{W}$ is directed along the $y$ axis,
\begin{equation}
\vec{W}\propto\vec{E}_a\times\vec{B}_a
=(0,E_aB_a,0),
\end{equation}
so that the wave propagates in the $y$ direction. Because the electromagnetic wave is confined between the two conducting slabs, it forms a standing wave.

The spatial dependence of the electric and magnetic fields can be written in the form
\begin{equation}
E_a(y)=-E_{a0}
\frac{\cos(m_a y+\delta)}
{\cos\delta},
\end{equation}
and
\begin{equation}
B_a(y)=E_{a0}
\frac{\sin(m_a y+\delta)}
{\cos\delta},
\end{equation}
where
\begin{equation}
E_{a0}=g_{a\gamma\gamma}a_0B_t
\end{equation}
is the amplitude of the axion-induced electric field in the absence of cavity enhancement. The phase $\delta$ is determined by the boundary conditions at the conducting slabs,
\begin{equation}
\tan\delta=
\frac{\sin(m_al)}
{1+\cos(m_al)}.
\end{equation}

The finite conductivity $\sigma$ of the conducting slabs allows the electromagnetic field to penetrate into the slabs over the skin depth
\begin{equation}
\delta_e=
\sqrt{\frac{2}{m_a\sigma}},
\end{equation}
where natural units are used. Near resonance, the cavity field is strongly enhanced. In particular, at the fundamental resonance,
$
l\simeq\frac{\pi}{m_a},
$
the electric and magnetic fields inside the cavity can be expressed approximately as
\begin{equation}
E_a(y)
\simeq
\sqrt{2}E_{a0}
\frac{\sin(m_ay)}
{m_a\delta_e},
\end{equation}
and
\begin{equation}
B_a(y)
\simeq
-\sqrt{2}E_{a0}
\frac{\cos(m_ay)}
{m_a\delta_e}.
\end{equation}
Thus, the cavity field is enhanced relative to the axion-induced field by a factor of order
\begin{equation}
\frac{1}{m_a\delta_e}.
\end{equation}
For a good conducting cavity wall, $m_a\delta_e\ll1$, and the resonant enhancement can consequently be very large. In the following discussion, the time dependence of the fields, which is simply harmonic at approximately the axion frequency, is suppressed for notational simplicity.

We next consider the orientation of the semiconductor sample relative to the external magnetic field. If the 2D electron layer were exactly perpendicular to the external magnetic field, the microwave electric field would be perpendicular to the 2D electron plane and would not efficiently drive in-plane electronic currents. We therefore tilt the 2D electron layer by a small angle so that the microwave electric field acquires a component parallel to the 2D electron system.

As shown in Fig.(\ref{b}), we take, for definiteness,
\begin{equation}
\theta=\frac{\pi}{6}.
\end{equation}
The component of the static magnetic field perpendicular to the 2D electron layer is then
\begin{equation}
B_{\perp}=B_t\cos\theta
\simeq0.86B_t,
\end{equation}
while the component parallel to the 2D electron layer is
\begin{equation}
B_{\parallel}=
B_t\sin\theta
=0.5B_t.
\end{equation}
The perpendicular component $B_\perp$ determines the filling factor and hence the quantum Hall state,
\begin{equation}
\nu=
\frac{2\pi\rho_e}{eB_{\perp}},
\end{equation}
where $\rho_e$ is the 2D electron density.

The electromagnetic field inside the semiconductor must also be taken into account because the semiconductor has a dielectric constant different from that of vacuum. For GaAs, we adopt the relative dielectric constant
$
\epsilon\simeq13,
$
while the relative permittivity of vacuum is taken to be unity in the present natural-unit convention.

The microwave wave is refracted at the semiconductor surface. For an incident angle
\begin{equation}
\theta_i=\frac{\pi}{2}-\theta
=\frac{\pi}{3},
\end{equation}
Snell's law gives
\begin{equation}
\sqrt{\epsilon}\sin\theta_r
=\sin\theta_i,
\end{equation}
and hence
\begin{equation}
\sin\theta_r\simeq0.24.
\end{equation}
Taking account of the electromagnetic boundary conditions at the semiconductor interface, the electric and magnetic field amplitudes inside the semiconductor are approximately
\begin{equation}
E'_a=
\frac{2\cos\theta_i}
{\sqrt{\epsilon}\cos\theta_i+\cos\theta_r}
E_a
\simeq0.36E_a,
\end{equation}
and
\begin{equation}
B'_a
\simeq
\sqrt{\epsilon}E'_a
\simeq1.3E_a.
\end{equation}
Here, the prime denotes electromagnetic fields inside the semiconductor.

The component of the electric field parallel to the 2D electron layer is therefore
\begin{equation}
E'_{\parallel}=
E'_a\cos\theta_r
\simeq0.97E'_a
\simeq0.35E_a.
\end{equation}

The microwave magnetic field is essentially parallel to the 2D electron layer in this geometry. Thus,
\begin{equation}
B'_a\simeq1.3E_a.
\end{equation}

The magnitude of the electromagnetic energy flux incident on the 2D electron system is consequently proportional to
\begin{equation}
E'_{\parallel} B'_a
\simeq0.46E_a^2.
\end{equation}
More precisely, the energy flux is given by the appropriate component of the Poynting vector $\vec{W}$, with the corresponding electromagnetic-unit convention understood. The above relation is sufficient for estimating the relative magnitude of the microwave field coupled to the 2D electron system.

The relevant field components can therefore be summarized as
\begin{equation}
E'_{\parallel}\simeq0.35E_a,
\qquad
B'\simeq1.3E_a,
\qquad
B_{\perp}\simeq0.86B_t,
\end{equation}
and
\begin{equation}
E'_{\parallel} B'_a\simeq0.46E_a^2.
\end{equation}

The in-plane electric field $E'_{\parallel}$ drives an electrical current in the 2D electron system. In the quantum Hall regime, the longitudinal and transverse components of the current are determined by the conductivity tensor,
Thus, the microwave electric field produces both a longitudinal current proportional to $\sigma_{xx}$ and a transverse Hall current proportional to $\sigma_{xy}$.

This distinction is crucial for the proposed axion detection scheme. In the metallic quantum Hall regime, $\mathrm{Re}(\sigma_{xx}(\omega))$ is finite, and the axion-induced microwave field is dissipated through the longitudinal current. The absorbed power is proportional to
\begin{equation}
P_{\rm abs}
\propto
\mathrm{Re}(\sigma_{xx}(\omega))
|E'_\parallel|^2.
\end{equation}
This dissipation loads the cavity and reduces its quality factor.

On a quantum Hall plateau, in contrast,
\begin{equation}
\mathrm{Re}(\sigma_{xx}(\omega))\simeq 0,
\quad
\sigma_{xy}\neq 0.
\end{equation}
The longitudinal dissipative response is therefore strongly suppressed, while the transverse Hall response remains finite. The axion-induced electric field can consequently generate a Hall current without producing a comparably large energy loss in the cavity. This allows the cavity to maintain a large loaded quality factor while providing a measurable electrical signal through the Hall current.

The essential feature of the proposed method is therefore the simultaneous realization of
\begin{equation}
\mathrm{Re}(\sigma_{xx}(\omega))\simeq 0,
\quad
\sigma_{xy}=\frac{\nu e^2}{2\pi}
\end{equation}
in the quantum Hall plateau state. The former suppresses microwave dissipation and preserves the cavity enhancement, whereas the latter provides the observable Hall-current signal.

\section{absorption by dissipative quantum Hall state}

We first consider a metallic quantum Hall state with a finite longitudinal conductivity,
\begin{equation}
\mathrm{Re}(\sigma_{xx}(\omega))\neq0,
\end{equation}
and estimate the amount of axion-induced microwave radiation that can be absorbed by the 2D electron system. In this regime, the quantum Hall sample acts essentially as a substitute for the metallic antenna conventionally used in axion haloscopes.

For a 2D electron system with surface area $S$, the time-averaged power dissipated by the longitudinal current is
\begin{equation}
P_s
=
\frac{1}{2}S
\mathrm{Re}\left(\sigma_{xx}(\omega)\right)
|E'_\parallel|^2,
\label{Ps_basic}
\end{equation}
where $E'_\parallel$ denotes the component of the microwave electric field parallel to the 2D electron layer. The factor $1/2$ arises from the time average of the square of a harmonic electric field.

In the absence of resonant enhancement in the cavity, the parallel electric field is approximately
\begin{equation}
E'_\parallel=0.35E_a,
\end{equation}
where
\begin{equation}
E_a=g_{a\gamma\gamma}a_0B_t,
\qquad
a_0=\frac{\sqrt{2\rho_d}}{m_a}.
\end{equation}
At resonance, however, the electric field in the cavity is enhanced by the cavity quality factor, and hence
\begin{equation}
E'\propto Q_L.
\end{equation}
The dissipated power is therefore strongly enhanced by the resonant cavity.

The longitudinal conductivity reaches its largest value near the center of a plateau-to-plateau transition. As a representative example, we consider a state near $\nu=3/2$, for which experimentally observed values of the longitudinal conductivity are of order
\begin{equation}
\mathrm{Re}(\sigma_{xx})
\simeq
\frac{0.2e^2}{h}
=
\frac{0.1e^2}{\pi}.
\end{equation}
Although it may be experimentally difficult to tune the magnetic field precisely to the point of maximum longitudinal conductivity, this value provides a useful estimate of the largest microwave absorption by the 2D electron system.

Using the electromagnetic field components obtained in the previous section, the energy flux incident on the 2D electron layer is proportional to
\begin{equation}
0.46SE_a^2.
\end{equation}
The fraction of this incident electromagnetic power dissipated by the 2D electron system is then
\begin{equation}
\frac{P_s}{0.46SE_a^2}
=
\frac{1}{2}
\frac{\mathrm{Re}(\sigma_{xx}(m_a))(0.35)^2}{0.46}
\simeq
4\times10^{-3}
\left(
\frac{\mathrm{Re}(\sigma_{xx}(m_a))}
{0.029}
\right),
\label{absorption_fraction}
\end{equation}
where
$
0.029\simeq\frac{0.1e^2}{\pi}.
$
Thus, even in the metallic quantum Hall state, only a small fraction of the incident microwave power is dissipated by the 2D electrons.

The relevant axion-induced microwave frequency is
\begin{equation}
\frac{m_a}{2\pi}
\simeq
2.4\mathrm{GHz}
\left(\frac{m_a}{10^{-5}\mathrm{eV}}\right).
\end{equation}
Experimental measurements indicate that the maximum value of
$\mathrm{Re}(\sigma_{xx}(\omega))$ in the transition region depends only weakly on frequency over the range of approximately $0.5$--$10$ GHz, as illustrated in Fig.(\ref{aa}) \cite{sigma,sigma2}. We therefore use the measured value above as a representative estimate at the axion frequency.

On a well-developed quantum Hall plateau, in contrast,
\begin{equation}
\mathrm{Re}(\sigma_{xx}(\omega))\simeq0,
\end{equation}
and the microwave power dissipated by the 2D electrons is correspondingly strongly suppressed. This should not be interpreted as implying that there is necessarily no electromagnetic reflection at the semiconductor interface; rather, the dissipative absorption associated with the longitudinal conductivity is negligible. In the present geometry, the intrinsic electromagnetic reflection associated with the very thin 2D electron layer is also expected to be small for the parameters considered here.

\vspace{0.1cm}

We next evaluate the effect of the semiconductor sample on a cylindrical resonant cavity. The power generated by axion dark matter in a haloscope, in the absence of additional losses, can be written as
\begin{equation}
P_0
=g_{a\gamma\gamma}^2B_t^2
\frac{\rho_d}{m_a}
CVQ_0,
\label{P0}
\end{equation}
where $V$ is the cavity volume, $C$ is the cavity form factor, and $Q_0$ is the unloaded quality factor of the cavity. For example, the ADMX experiment uses a form factor of order $C\sim0.4$ for its $\mathrm{TM}_{010}$ mode \cite{admx1,admx2}. In the present analysis, we take
\begin{equation}
C=0.6.
\end{equation}
For an ideal cylindrical $\mathrm{TM}_{010}$ mode without additional perturbations, the form factor can be as large as approximately $0.7$.

The unloaded quality factor $Q_0$ is determined primarily by the electromagnetic loss in the cavity walls. For the cylindrical cavity considered here, we take
\begin{equation}
Q_0\sim10^5-10^6,
\end{equation}
depending on the conductivity and material of the cavity wall. In the following, we adopt the conservative value
\begin{equation}
Q_0=10^5,
\end{equation}
which is smaller than the effective quality factor of the axion field,
$
Q_a\simeq10^6.
$

When a semiconductor sample containing a 2D electron system is inserted into the cavity, the sample introduces an additional electromagnetic loss. We characterize this loss by a sample quality factor $Q_s$, defined by
\begin{equation}
Q_s
=\frac{\omega U}{P_s}
\simeq
\frac{m_aU}{P_s},
\label{Qs_definition}
\end{equation}
where $U$ is the electromagnetic energy stored in the cavity. The loaded cavity quality factor is consequently reduced by the additional sample loss. We write
\begin{equation}
\frac{1}{Q_L}
=
\frac{1}{Q_0}
+
\frac{1}{Q_s},
\label{QL}
\end{equation}
with the finite axion linewidth treated separately when evaluating the signal bandwidth.

For an axion field resonant with the cavity, the axion-induced power deposited in the cavity is therefore
\begin{equation}
P_L
=g_{a\gamma\gamma}^2B_t^2
\frac{\rho_d}{m_a}
CVQ_L.
\label{PL}
\end{equation}
The corresponding average electric field in the cavity follows from
\begin{equation}
Q_LP_L=m_aU,
\end{equation}
together with
\begin{equation}
U\simeq\frac{VE_c^2}{2}.
\end{equation}
We then obtain
\begin{equation}
E_c=
\sqrt{2}
g_{a\gamma\gamma}B_t
\frac{\sqrt{C\rho_d}}{m_a}
Q_L.
\label{11}
\end{equation}
Thus, the electric field in the cavity is proportional to $Q_L$, as expected for resonant enhancement.

The electric field parallel to the 2D electron layer is
\begin{equation}
E'_\parallel=0.35E_c.
\end{equation}
Combining Eqs.~(\ref{Ps_basic}), (\ref{Qs_definition}), and (\ref{11}), we obtain
\begin{equation}
Q_s
=\frac{Vm_a}
{1.2\times10^{-1}S
\mathrm{Re}[\sigma_{xx}(m_a)]}.
\label{Q_s}
\end{equation}
For the cylindrical cavity at resonance,
\begin{equation}
R=\frac{2.4}{m_a}
\simeq
4.8\mathrm{cm}
\left(
\frac{10^{-5}\mathrm{eV}}{m_a}
\right),
\end{equation}
and hence
\begin{equation}
V=\pi R^2l_{\rm cav}
\simeq
7.2\mathrm{l}
\left(
\frac{10^{-5}\mathrm{eV}}{m_a}
\right)^2
\left(
\frac{l_{\rm cav}}{100\mathrm{cm}}
\right).
\end{equation}
Using
$
\mathrm{Re}(\sigma_{xx}(m_a))
\simeq
\frac{0.1e^2}{\pi},
$
as suggested by measurements at microwave frequencies, we obtain
\begin{equation}
Q_s\simeq
1.1\times10^6
\left(
\frac{10\mathrm{cm}^2}{S}
\right)
\left(
\frac{0.1e^2/\pi}
{\mathrm{Re}(\sigma_{xx})}
\right)
\left(
\frac{10^{-5}\mathrm{eV}}{m_a}
\right).
\label{Qs_final}
\end{equation}

The relatively large value of $Q_s$ reflects the weak microwave dissipation in the thin 2D electron system. The corresponding cavity coupling parameter is
$
\beta\equiv\frac{Q_0}{Q_s}\simeq 0.091,
$
which is smaller than unity, for the parameters considered here. Thus, the semiconductor sample weakly loads the cavity. For
\begin{equation}
Q_0=10^5,
\qquad
Q_s\simeq1.1\times10^6,
\end{equation}
the loaded quality factor is approximately
\begin{equation}
Q_L\simeq9.2\times10^4.
\end{equation}

\vspace{0.1cm}

We can now estimate the power dissipated by the 2D electrons in the metallic quantum Hall state. Since the fraction of the cavity power dissipated by the sample is determined by the ratio of the corresponding quality factors, we have
\begin{equation}
P_s=
P_L\frac{Q_L}{Q_s}.
\end{equation}
Consequently,
\begin{equation}
P_s(S=10\mathrm{cm}^2)=
\left(g_{a\gamma\gamma}B_t\right)^2
\frac{\rho_d}{m_a}
VC
\frac{Q_L^2}{Q_s},
\end{equation}
or numerically,
\begin{equation}
\label{P}
P_s(S=10\mathrm{cm}^2)
\simeq
6.2\times10^{-24}\mathrm{W}
\left(\frac{g_\gamma}{0.36}\right)^2
\left(\frac{B_t}{1.5\times 10^5\mathrm{G}}\right)^2
\left(\frac{V}{7.2\mathrm{l}}\right)
\left(\frac{C}{0.6}\right)
\left(\frac{\rho_d}{0.45\mathrm{GeVcm^{-3}}}\right).
\end{equation}
Here, the metallic quantum Hall state acts as a dissipative microwave absorber, analogous to a conventional metallic antenna.

An interesting feature of this result is that, under the assumptions made above, the absorbed power is approximately independent of the axion mass. Indeed,
\begin{equation}
g_{a\gamma\gamma}^2\propto m_a^2,
\qquad
V\propto m_a^{-2},
\qquad
Q_s\propto m_a^{-1},
\end{equation}
and therefore
\begin{equation}
\frac{g_{a\gamma\gamma}^2V}{m_aQ_s}
\simeq\mathrm{constant}.
\end{equation}
This scaling assumes that the 2D electron area $S$ is kept independent of $m_a$. Such an assumption is possible provided that the physical dimensions of the semiconductor sample remain smaller than the cavity dimensions, whose characteristic radius scales as
\begin{equation}
R\propto m_a^{-1}.
\end{equation}

\vspace{0.1cm}

We next compare the absorbed power with the thermal noise power in the relevant bandwidth. Taking the axion signal bandwidth to be
$
\delta\omega\simeq10^{-6}m_a,
$
the thermal noise power is approximately
\begin{equation}
P_n
=\frac{T\delta\omega}{2\pi}
\simeq
3.3\times10^{-21}\mathrm{W}
\left(\frac{T}{100\mathrm{mK}}\right)
\left(\frac{m_a}{10^{-5}\mathrm{eV}}\right).
\end{equation}

For an observation time $\delta t_{\rm ob}$, the corresponding radiometer-type signal-to-noise ratio is estimated as
\begin{equation}
\frac{P_s}{P_n}
\sqrt{\frac{\delta\omega \delta t_{\rm ob}}{2\pi}}
\simeq
0.92
\left(\frac{g_\gamma}{0.36}\right)^2
\left(\frac{B_t}{1.5\times 10^5\mathrm{G}}\right)^2
\left(\frac{100\mathrm{mK}}{T}\right)
\left(\frac{V}{7.2\mathrm{l}}\right)
\left(\frac{10^{-5}\mathrm{eV}}{m_a}\right)^{1/2}
\left(\frac{C}{0.6}\right)
\left(\frac{\rho_d}{0.45\mathrm{GeVcm^{-3}}}\right)
\left(\sqrt{\frac{\delta t_{\rm ob}}{100\mathrm{s}}}\right).
\label{SNR}
\end{equation}
For $\delta t_{\rm ob}=100\mathrm{s}$,
\begin{equation}
\sqrt{\frac{\delta\omega \delta t_{\rm ob}}{2\pi}}
\simeq
4.9\times10^2
\sqrt{\frac{\delta t_{\rm ob}}{100\mathrm{s}}}.
\end{equation}
Thus, the dissipative quantum Hall state can provide a signal at a level approaching the thermal-noise background for the parameters adopted here.

\vspace{0.1cm}

For comparison with the Hall-current signal discussed in the next section, we also estimate the longitudinal current in the metallic quantum Hall state. The current is
\begin{equation}
I_{xx}=
\sigma_{xx}E'_\parallel l,
\end{equation}
where $l$ is the transverse dimension of the 2D electron system. Taking
$
\mathrm{Re}(\sigma_{xx})=
\frac{0.1e^2}{\pi}
$
near the center of the transition region, we obtain
\begin{equation}
I_{xx}
\simeq
2.2\times10^{-19}\mathrm{A}
\left(\frac{g_\gamma}{0.36}\right)
\sqrt{\frac{C}{0.6}}
Q_L
\left(\frac{\rho_e}{3\times10^{11}\mathrm{cm}^{-2}}\right)
\left(\frac{l}{10\mathrm{cm}}\right).
\end{equation}
Here, we have used
\begin{equation}
\frac{2\pi\rho_e}{eB_{\perp}}=\frac32,
\end{equation}
which gives
\begin{equation}
B_{\perp}\simeq8\times10^4\mathrm{G}.
\end{equation}

\vspace{0.1cm}

The metallic quantum Hall state can therefore serve as a substitute for the metallic antenna used in a conventional haloscope. In standard haloscopes, the coupling between the cavity and the antenna is usually adjusted through the antenna insertion depth. In terms of the coupling parameter
$
\beta=\frac{Q_0}{Q_s},
$
the coupling can thus be optimized experimentally to obtain an appropriate balance between cavity enhancement and signal extraction.

For a quantum Hall sample, however, the dissipative quality factor $Q_s$ is primarily determined by the sample area, its longitudinal conductivity, and its electromagnetic coupling to the cavity. Unlike a conventional movable antenna, there is no equally simple insertion-depth parameter with which to continuously tune the coupling. Consequently, optimization of the coupling is less flexible in the metallic quantum Hall configuration. This provides an additional reason to consider the insulating quantum Hall state, where the suppression of $\mathrm{Re}(\sigma_{xx})$ minimizes the additional cavity loss while the finite Hall conductivity can still provide a measurable signal.

\section{Hall current in nondissipative quantum Hall state}

The metallic quantum Hall state discussed in the previous section does not provide a fundamental advantage over a conventional metallic antenna. In the transition region between adjacent quantum Hall plateaus, the longitudinal conductivity $\mathrm{Re}(\sigma_{xx})$ has a pronounced maximum, as shown in Fig.(\ref{aa}). Consequently, it is necessary to tune the magnetic field to a relatively narrow range in order to realize a state with a large longitudinal conductivity,
\begin{equation}
\mathrm{Re}(\sigma_{xx})\sim\frac{0.1e^2}{\pi}.
\end{equation}
Moreover, the resulting microwave dissipation loads the cavity and reduces its quality factor. A conventional antenna can instead be mechanically coupled to the cavity and its coupling strength optimized by adjusting the antenna insertion depth. Thus, a conventional antenna is generally more effective than a metallic quantum Hall sample for detecting the dissipated microwave power.

\vspace{0.1cm}
A qualitatively different situation arises when the quantum Hall system is operated on a well-developed insulating plateau, where
\begin{equation}
\sigma_{xy}=\frac{\nu e^2}{2\pi}
\quad \mbox{and} \quad
\mathrm{Re}(\sigma_{xx})\simeq0.
\end{equation}
In this regime, a large Hall current can be generated by the microwave electric field while the associated energy dissipation remains negligibly small.

To see this explicitly, the current density in a quantum Hall system is given by
the longitudinal component,
\begin{equation}
I_{xx}=\mathrm{Re}(\sigma_{xx})E'_\parallel\simeq 0,
\end{equation}
which is parallel to the electric field and therefore the dissipation of electromagnetic energy is strongly suppressed. 
On the other hand, the Hall component of 
induced current,
\begin{equation}
 I_H
=
\mathrm{Re}(\sigma_{xy})E'_\parallel,
\end{equation}
is perpendicular to the electric field. Consequently,
\begin{equation}
\vec{I}_H\cdot \vec{E}_\parallel'=0,
\end{equation}
and the Hall current does not dissipate energy in the ideal quantum Hall state. The dissipated power is instead determined by the longitudinal response,
\begin{equation}
P_{\rm diss}
\propto
\mathrm{Re}(\sigma_{xx}(\omega))|E'_\parallel|^2\simeq 0.
\end{equation}

This property is absent in a conventional metallic antenna. A current flowing in a metallic antenna necessarily encounters a finite dissipative resistance and therefore extracts energy from the microwave field. In contrast, the quantum Hall state can simultaneously support a finite transverse current and a vanishing longitudinal dissipative response. This provides the essential advantage of the proposed detection method.

\vspace{0.1cm}
Because the additional dissipative loss of the quantum Hall sample is strongly suppressed on a plateau, the loaded cavity quality factor can approach the unloaded value,
\begin{equation}
Q_L\simeq Q_0.
\end{equation}
The axion-induced electric field is consequently enhanced by a large factor of order $Q_L$. The Hall current generated by this enhanced electric field is therefore
\begin{equation}
I_H
=
\mathrm{Re}(\sigma_{xy})E'_\parallel l,
\end{equation}
where $l$ is the dimension of the 2D electron system perpendicular to the Hall-current direction.
Hereafter, we write $\mathrm{Re}(\sigma_{xy}(m_a))$ simply as $\sigma_{xy}$.  
Using
\begin{equation}
E'_\parallel=0.35E_c
=
0.35\sqrt{C}Q_L E_a,
\end{equation}
and
\begin{equation}
E_a=g_{a\gamma\gamma}a_0B_t,
\end{equation}
we obtain
\begin{equation}
I_H
=
\sigma_{xy}(0.35)g_{a\gamma\gamma}a_0B_t
\sqrt{C}Q_Ll
=
\left(\frac{\nu e^2}{2\pi}\right)
(0.35)g_{a\gamma\gamma}a_0B_t
\sqrt{C}Q_Ll ,
\label{Ih}
\end{equation}
with $\sigma_{xy}\simeq \nu e^2/2\pi$.

It leads to 
\begin{equation}
\label{I_H}
I_H
\simeq
1.66\times10^{-18}\mathrm{A}
\left(\frac{g_\gamma}{0.36}\right)
\sqrt{\frac{C}{0.6}}Q_L
\left(\frac{\rho_e}{3\times10^{11}\mathrm{cm}^{-2}}\right)
\left(\frac{l}{10\mathrm{cm}}\right),
\end{equation}
where we have used a 2D electron density
\begin{equation}
\rho_e=3\times10^{11}\mathrm{cm}^{-2}.
\end{equation}
The formula in eq(\ref{I_H}) is independent of the filling factor $\nu$. When we choose a quantum Hall state with a filling factor $\nu$,
 corresponding perpendicular magnetic field realizing the state is determined from
$
\frac{2\pi\rho_e}{eB_{\perp}}=\nu,
$
giving approximately
\begin{equation}
B_{\perp}\simeq1.2\times10^5\mathrm{G}\Big(\frac{1}{\nu}\Big)\left(\frac{\rho_e}
{3\times10^{11}\mathrm{cm}^{-2}}\right).
\end{equation}

Since the sample is tilted by $\theta=\pi/6$,
$
B_\perp=0.86B_t
$
and therefore external magnetic field we imposed is
\begin{equation}
B_t\simeq1.4\times10^5\mathrm{G}\Big(\frac{1}{\nu}\Big)\left(\frac{\rho_e}
{3\times10^{11}\mathrm{cm}^{-2}}\right).
\end{equation}

As is evident, increasing both the density $\rho_e$ and the filling factor $\nu$ by the same factor yields a larger Hall current, even though the magnetic field strength remains the same. It is a specific feature of the quantum Hall system.

\vspace{0.1cm}

We note that the Hall current $I_H\simeq 1.66\times10^{-18}\mathrm{A}$ at $\nu=2$ 
realized with $B_{\perp}\simeq 6\times 10^4\mathrm{G} $ is bigger than the current $I_{xx}\simeq 2.2\times10^{-19}\mathrm{A}$ at $\nu=3/2$
realized with $ B_{\perp}\simeq 8\times 10^4\mathrm{G} $. Thus, we find 
that the use of nondissipative quantum Hall state on plateau of $\nu=2$
for the axion detection is more favorable than the dissipative quantum Hall state with $\nu=3/2$.

The equality
$
\frac{2\pi\rho_e}{eB_\perp}=\nu
$
specifies the center of the $\nu$ plateau. Importantly, this equality need not hold exactly throughout the plateau. Once the Fermi energy lies in localized states, the Hall conductivity remains approximately quantized,
\begin{equation}
\sigma_{xy}\simeq\frac{\nu e^2}{2\pi},
\end{equation}
over a finite range of magnetic field. Therefore, the Hall-current estimate in Eq.(\ref{I_H}) remains valid throughout the plateau, provided that the Hall conductivity remains well quantized. In the sense, it is favorable to have quantum Hall state with large plateau.
It is realized in the states with the filling factor $\nu=2$, or $\nu=4$.

\vspace{0.1cm}
Supposing a 2D electron area $S=10^{2}\mathrm{cm}^2$, for instance
\begin{equation}
S\simeq10\mathrm{cm}^2
\left(\frac{l}{10\mathrm{cm}}\right)
\left(\frac{L}{1\mathrm{cm}}\right),
\end{equation}
with the dimension $L$ along the Hall-current direction, 
we estimate the signal-to-noise ratio of the Hall-current measurement. The relevant thermal current noise can be expressed in terms of the Hall resistance,
\begin{equation}
R_s
=
\rho_{xy}\frac{L}{l}
=
\frac{L}{\sigma_{xy}l}\propto \nu^{-1}.
\end{equation}
For a bandwidth $\delta\omega=10^{-6}m_a$, the corresponding Johnson--Nyquist current noise is
\begin{equation}
I_n
=
\sqrt{\frac{2T\delta\omega}{\pi R_s}}
=
\sqrt{
\frac{2T\delta\omega \sigma_{xy}l}
{\pi L}
}\propto \nu^{1/2}.
\end{equation}
The ratio of the Hall-current signal to the current noise is therefore
\begin{equation}
\frac{I_H^2}{I_n^2}=
\frac{\pi\sigma_{xy}E_{\parallel}'^2lL}{2T\delta \omega}=\frac{\pi\sigma_{xy}E_{\parallel}'^2S}
{2T\delta\omega}.
\label{INratio}
\end{equation}
Using
\begin{equation}
\sigma_{xy}=\frac{\nu e^2}{2\pi},
\qquad
E'_\parallel=0.35\sqrt{C}Q_LE_a,
\end{equation}
and
\begin{equation}
B_t=\frac{B_\perp}{0.86},
\qquad
B_\perp=\frac{2\pi\rho_e}{e\nu},
\end{equation}
we find
\begin{equation}
\label{R}
\frac{I_H^2}{I_n^2}
\propto
\frac{\rho_e^2}{\nu}
\frac{(g_{a\gamma\gamma}a_0)^2
C Q_L^2lL}
{T\delta\omega}.
\end{equation}
Thus, for fixed 2D electron density, sample dimensions, temperature, cavity quality factor, and axion bandwidth,
\begin{equation}
\frac{I_H^2}{I_n^2}\propto\frac{1}{\nu}.
\end{equation}

The origin of this dependence is transparent. The Hall conductivity decreases linearly with $\nu$, but the magnetic field required to realize a given filling factor increases as $1/\nu$. Thus, the Hall current $I_H\propto \sigma_{xy}B_t$ is independent of $\nu$. 
But, the noise current is 
such that $I_n\propto \nu^{1/2}$. Therefore, we have the ratio $I_H^2/I_n^2\propto 1/\nu$.

\vspace{0.1cm}
It is important to note that the ratio $I_H^2/I_n^2\propto S=lL$. That is, the ratio increases as the surface area of 2D electrons
increases. It is a specific feature of the quantum Hall state.
In the case of metal antenna, the ratio $P/P_n\propto B^2$ increases as the magnetic field $B$ increases. 
It is implicit in the quantum Hall state because $I_H^2/I_n^2\propto \rho_e^2\propto B^2$. But, there is 
no such a property in the metal antenna that $I_H^2/I_n^2\propto S$. The increase of $S$ in the metal antenna leads to 
large dissipation energy so that the lorded quality factor $Q_L$ decreases and the power $P$ decreases.
However, we may increase the $S/N$ ratio by making $S$ large without decreasing the quality factor $Q_L$ in
the quantum Hall state.

\vspace{0.1cm}
The center of a quantum Hall plateau occurs approximately at
\begin{equation}
B_\perp
=
\frac{2\pi\rho_e}{e\nu}
\simeq
1.2\times10^5\mathrm{G}
\left(\frac{1}{\nu}\right)
\left(\frac{\rho_e}
{3\times10^{11}\mathrm{cm}^{-2}}\right).
\end{equation}
Thus, for a semiconductor sample with fixed $\rho_e$, different quantum Hall plateaus can be explored simply by varying the applied magnetic field. The $\nu=1/3$ plateau is particularly attractive from the viewpoint of the Hall-current signal because the estimated $S/N$ increases as $\nu$ decreases.

Nevertheless, the smallest accessible filling factor is not necessarily optimal experimentally. Fractional quantum Hall plateaus generally become narrower and more sensitive to temperature and disorder as the filling factor decreases. Furthermore, to realized the state
with small filling factor $\nu=1/3$, we need much large magnetic field $\sim 3.6\times 10^5\mathrm{G}$.
The choice of $\nu=1/3$ therefore does not represent a compromise between a favorable Hall-current signal and the experimental accessibility of the quantum Hall state.

\vspace{0.1cm}

Rather, we consider the integer quantum Hall state at $\nu=2$. This state is generally easier to realize and can be more robust against disorder and temperature than the fractional $\nu=1/3$ state. Moreover, the excitation gap of the integer quantum Hall state can be larger than that of the fractional state. It is therefore reasonable to expect a robust Hall response at microwave frequencies provided that the photon energy remains sufficiently smaller than the relevant excitation gap.

For the axion mass considered here,
$
m_a=10^{-5}\mathrm{eV},\,\,
\frac{m_a}{2\pi}\simeq2.4,\mathrm{GHz},
$
the microwave photon energy is below the characteristic excitation energy of a well-developed integer quantum Hall state. We therefore expect the quantized Hall response to remain approximately intact,
\begin{equation}
\mathrm{Re}(\sigma_{xy}(\omega=m_a))
\simeq
\sigma_{xy}(0)
=
\frac{e^2}{2\pi},
\end{equation}
although this frequency dependence should ultimately be verified experimentally.

For $\nu=2$, the estimated Hall-current signal-to-noise ratio remains large. Using the same parameters as above, we obtain

\begin{eqnarray}
\label{con}
\frac{I_H^2}{I_n^2}\sqrt{\frac{\delta \omega \delta t}{2\pi}}&\simeq& 1.38\Big(\frac{g_{\gamma}}{0.36}\Big)^2 
\Big(\frac{C}{0.6}\Big)\Big(\frac{Q_L}{10^5}\Big)^2\Big(\frac{\rho_e}{3\times 10^{11}\mathrm{cm}^{-2}}\Big)^2\nonumber \\
&\times &\Big(\frac{100\mathrm{mK}}{\mathrm{T}}\Big)\Big(\frac{10^{-5}\mathrm{eV}}{m_a}\Big)^{1/2} 
\Big(\frac{\rho_d}{0.45\mathrm{GeVcm^{-3}}}\Big)\Big(\frac{S}{10\mathrm{cm}^2}\Big)
\sqrt{\frac{\delta t}{10^2\mathrm{s}}}
\end{eqnarray}
for an observation time $\delta t=100\mathrm{s}$ under the assumed experimental conditions. Thus, although the $\nu=1/3$ state provides a larger $S/N$ according to the scaling above, the integer quantum Hall state with $\nu=2$ may be experimentally more advantageous because of its greater robustness and simpler realization. The state is realized with magnetic field 
$B_t\simeq 7\times 10^4\mathrm{G}$.

\vspace{0.1cm}
We should note a following scaling derived from eq(\ref{R}). 
When we scale $\rho_e$ and $\nu$ as $\rho_e\to \lambda \rho_e$ and $\nu \to \lambda \nu$,
we have $S/N\equiv I_H^2(\sqrt{\delta \omega \delta t/2\pi})/I_n^2 \to \lambda S/N$, while magnetic field $B\propto \rho_e/\nu$
remain unchanged.
Therefore, 
when we use the integer quantum Hall state 
with electron density $\rho_e=6\times 10^{11}\mathrm{cm}^2$ and $\nu=4$, the signal-to-noise ratio becomes large $\simeq 2.8$,
but magnetic field $B_t\simeq 7\times 10^4\mathrm{G}$ remains unchanged.
Thus, the state with $\nu=4$ and $\rho_e=6\times 10^{11}\mathrm{cm}^2$ is more favorable for the axion detection. 

\vspace{0.1cm}

Finally, the angle between the 2D electron layer and the applied magnetic field provides an additional experimental parameter for optimizing the Hall-current signal. In the geometry considered above,
\begin{equation}
B_\perp=B_t\cos\theta,
\end{equation}
while the electric field component parallel to the 2D electron layer depends on the refraction geometry and hence on $\theta$. A change in $\theta$ therefore changes both the perpendicular magnetic field that determines the quantum Hall filling factor and the in-plane microwave electric field that drives the Hall current.

Within a sufficiently broad quantum Hall plateau, a moderate change in $B_\perp$ does not significantly alter the quantized Hall conductivity. At the same time, an increase in the in-plane electric-field component $E'_\parallel$ directly increases the Hall current,
\begin{equation}
I_H=\sigma_{xy}E'_\parallel l.
\end{equation}
The sample orientation can therefore be used as an additional experimental parameter to optimize the Hall-current signal while keeping the system within a well-developed quantum Hall plateau.

The essential principle of the proposed method is thus the use of a quantum Hall plateau as a nearly lossless microwave detector: the longitudinal dissipative response is suppressed, preserving a large cavity quality factor, while the finite quantized Hall conductivity converts the axion-induced microwave electric field into a measurable transverse current.

\section{conclusion}

We have proposed a new method for axion detection that exploits the quantum Hall effect in a haloscope. By tuning the dimensions of the haloscope to the axion frequency, the axion-induced microwave electric field is resonantly enhanced by the cavity quality factor \(Q_L\). The enhanced microwave field drives a transverse Hall current in a quantum Hall sample.

The essential feature of the proposed method is that the Hall current can be generated without significant energy dissipation. In an insulating quantum Hall state, the longitudinal dissipative response is strongly suppressed, \(\mathrm{Re}\,\sigma_{xx}\simeq0\), while the Hall conductivity remains finite. In particular, for the integer quantum Hall state, the microwave-induced Hall current is essentially nondissipative because the Hall current flows perpendicular to the applied electric field. Consequently, the quantum Hall sample introduces only a small additional loss to the cavity, and the loaded quality factor can remain close to the unloaded value, \(Q_L\simeq Q_0\), even in the presence of a sizable Hall current. This situation is fundamentally different from that of a conventional metallic antenna, in which the induced current is accompanied by Joule dissipation and therefore reduces the cavity quality factor.

The high quality factor consequently produces a large resonantly enhanced electric field inside the cavity, resulting in a Hall current proportional to the enhancement factor,

$$
I_H\propto Q_L.
$$

For a two-dimensional electron density of
\(\rho_e=3\times10^{11}\,\mathrm{cm}^{-2}\), we find that the signal-to-noise ratio for the integer quantum Hall state with \(\nu=2\) can be expressed as
\begin{eqnarray}
\frac{I_H^2}{I_n^2}
\sqrt{\frac{\delta\omega\delta t}{2\pi}}
&\simeq&
1.4
\left(\frac{g_{\gamma}}{0.36}\right)^2
\left(\frac{C}{0.6}\right)
\left(\frac{Q_L}{10^5}\right)^2
\left(\frac{\rho_e}{3\times10^{11}\mathrm{cm}^{-2}}\right)^2
\left(\frac{100\mathrm{mK}}{T}\right)
\left(\frac{10^{-5}\mathrm{eV}}{m_a}\right)^{1/2} \nonumber \\
&\times &\left(\frac{\rho_d}{0.45\mathrm{GeVcm}^{-3}}\right)
\left(\frac{S}{10\mathrm{cm}^2}\right)
\sqrt{\frac{\delta t}{10^2\mathrm{s}}},
\label{con}
\end{eqnarray}
where an observation time of \(\delta t_{\rm ob}=100\) s has been assumed. 
For the electron density $3\times 10^{11}\mathrm{cm}^2$,
the \(\nu=2\) integer quantum Hall state can be realized with an external magnetic field of approximately \(B_t\simeq 7\times10^4\) G.
When we use the integer quantum Hall state 
with $\nu=4$ and $\rho_e=6\times 10^{11}\mathrm{cm}^2$, the signal-to-noise ratio becomes large $\simeq 2.8$,
but with identical magnetic field $B_t\simeq 7\times 10^4\mathrm{G}$.
It should be noted that the ratio increases as the surface area $S$ of 2D electrons increases. It is a
property which is absent in the metal antenna. 

We have also considered the fractional quantum Hall state with \(\nu=1/3\). At fixed electron density, this state can provide a larger signal-to-noise ratio than the \(\nu=2\) state because the required magnetic field is higher and the resulting Hall response leads to a favorable scaling of the signal-to-noise ratio. However, realizing a robust \(\nu=1/3\) state generally requires higher-mobility samples and a stronger magnetic field, approximately \(B_t\simeq 3\times10^5\) G for the parameters considered here. Thus, from a practical experimental standpoint, the integer quantum Hall state with \(\nu \ge 2\) appears to be the more favorable choice.

In conclusion, the proposed detection scheme exploits a distinctive property of the quantum Hall state: a large transverse current can be generated while the longitudinal dissipative response remains strongly suppressed. This allows the cavity to retain a high quality factor while providing a measurable Hall-current signal. The integer quantum Hall state therefore offers a promising alternative to conventional metallic antennas for axion haloscope experiments, particularly in frequency and axion-mass regions where maintaining a high cavity quality factor is essential.


\begin{thebibliography}{99}
\bibitem{axion1}R. D. Peccei and H. R. Quinn, Phys. Rev. Lett. 38 (1977) 1440.
\bibitem{axion2}S. Weinberg, Phys. Rev. Lett. 40 (1978) 223.
\bibitem{axion3}F. Wilczek, Phys. Rev. Lett. 40 (1978) 279.
\bibitem{review}Maurizio Giannotti, arXiv: 2412.08733.
\bibitem{girvin}R. E. Prange and S. M. Girvin, Springer Verlag, New York(1987).
\bibitem{sikivie}P. Sikivie, Phys. Rev. D32 (1985) 2988.
\bibitem{admx1}C. Goodman, et.al.,  Phys. Rev. Lett. 134 (2025) 111002. 
\bibitem{admx2}T. Braine, et. al., Phys. Rev. Lett. 124, (2020) 101303.
\bibitem{dfsz}M. Dine, W. Fischler and M. Srednicki, Phys. Lett. 104B (1981) 199. 
\bibitem{dfsz1} A. R. Zhitnitsky, Sov. J. Nucl. Phys. 31 (1980) 260.
\bibitem{ksvz}J. E. Kim, Phys. Rev. Lett. 43, (1979) 103. 
\bibitem{ksvz1}M. A. Shifman, A. I. Vainshtein and V. I. Zakharov, Nucl. Phys. B166 (1980) 493.
\bibitem{localization}P. W. Anderson, Phys. Rev. 109 (1958) 1492.
\bibitem{andouemura}T. Ando and Y. Uemura, J. Phys. Soc. Jpn. 36 (1974) 959.
\bibitem{aoki1}H. Aoki and T. Ando, Phys. Rev. Lett. 54 (1985) 831.
\bibitem{aoki2}H. Aoki and T. Ando, J. Phys. Soc. Jpn. 54 ( 1985) 2238.
\bibitem{topology1}R. B. Laughlin, Phys. Rev. B 23 (1981) 5632.
\bibitem{topology2} D. J. Thouless, M. Kohmoto, P. Nightingale, and M. den Nijs, Phys. Rev. Lett. 49 (1982) 405.
\bibitem{sigma}L. W. Engel, D. Shahar, Q. Kurdak, and D. C. Tsui, Phys. Rev. Letts. 71 (1993) 2638.
\bibitem{sigma2}F. Hohls, U. Zeitler, R. J. Haug, R. Meisels, K. Dybko, and F. Kuchar,  Phys. Rev. Lett.  89 (2002) 276801.
\bibitem{joseph1} X. G. Wen and A. Zee, Phy. Rev. B 47, (1993) 2265.
\bibitem{joseph2} Z.F. Ezawa, and A. Iwazaki, Phys. Rev. B 47 (1993) 7295.
\bibitem{joseph3}I. B. Spielman, J. P. Eisenstein, L. N. Pfeiffer, and K. W. West, Phy. Rev. Lett, 84 (2000) 5808.
\bibitem{iwazaki}Z. F. Ezawa and A. Iwazaki, Phys. Rev. Lett. 70 (1993) 3119. 
\bibitem{iwazaki01}A. Iwazaki, PTEP 2022 (2022) 2, 021B01.
\bibitem{Wil}J. Preskill, M. B. Wise and F. Wilczek, Phys. Lett. 120B (1983) 127.
\bibitem{Wil1}L. F. Abbott and P. Sikivie, Phys. Lett. B120 (1983) 133.
\bibitem{Wil2}M. Dine and W. Fischler, Phys. Lett. B120 (1983) 137.


\end{thebibliography}
\end{document}